\documentclass[letter]{aa}

\newcommand{\bea}{\begin{eqnarray}}
\newcommand{\eea}{\end{eqnarray}}

\usepackage{orcidlink}
\usepackage{natbib}

\usepackage{url}
\usepackage{soul}
\usepackage{multirow} 
\usepackage{makecell} 
\usepackage{booktabs}

\usepackage{txfonts}
\usepackage{graphicx}
\usepackage{txfonts,textcomp}

\usepackage{natbib,twoopt}
\bibpunct{(}{)}{;}{a}{}{,}
\usepackage[hyphenbreaks]{breakurl}
\usepackage{hyperref}

\hypersetup{
  colorlinks,
  citecolor=cyan,
  linkcolor=magenta,
  urlcolor=teal,
}

\begin{document} 

\title{Volatile enrichment in low-mass planets as signatures of past planetary disruption}

\author{
    Mario Sucerquia\inst{1}\thanks{E-mail: \href{mailto:mario.sucerquia@univ-grenoble-alpes.fr}{mario.sucerquia@univ-grenoble-alpes.fr}}\orcidlink{0000-0002-8065-4199}
    \and
    Matías Montesinos\inst{2}\thanks{E-mail: \href{mailto:matias.montesinosa@usm.cl}{matias.montesinosa@usm.cl}}\orcidlink{0000-0001-9789-5098}
    \and
    Ana M. Agudelo\inst{3}\orcidlink{0009-0001-2706-9943}
    \and
    Nicolás Cuello\inst{1}\orcidlink{0000-0003-3713-8073}
}

\institute{
    Univ. Grenoble Alpes, CNRS, IPAG, 38000 Grenoble, France
    \and
    Departamento de Física, Universidad Técnica Federico Santa María, Avenida España 1680, Valparaíso, Chile
        \and
    Instituto de F\'isica - FCEN, Universidad de Antioquia, Calle 70 No. 52-21, Medell\'in, Colombia.
}

   \date{\today}

\abstract
{Tidal disruption and engulfment events around main-sequence stars -- such as the luminous red nova ZTF SLRN-2020 (a candidate planetary-engulfment event) -- reveal the destruction of close-in giant planets. While current observations focus on stellar accretion and inner dust emission, the fate of the volatile-rich material expelled during disruption remains poorly understood.}
{We investigate whether the H/He-rich gas expelled from the disrupted planet’s envelope and atmosphere can escape the inner system and be gravitationally captured by a low-mass outer planet (volatile-enriched planet (VEP)), potentially forming a transient atmosphere and producing detectable volatile contamination.}
{We model the outward diffusion of gas from a tidally stripped giant using 2D hydrodynamical simulations with \texttt{FARGO3D}, complemented by analytical estimates of volatile observability and atmospheric escape. We assess the efficiency of gas capture by outer planets and the survival timescales of the resulting secondary atmospheres under XUV-driven erosion.}
{Our simulations show that volatile-rich gas can form a VEP. The resulting envelopes can contain $10^{-10}$–$10^{-6},M_\oplus$ -- up to the mass of Earth's atmosphere -- for Earth-like planets, yielding transit depths of tens to hundreds of parts per million. Such signatures may persist for $10^6$–$10^8$ years, depending on planetary mass, orbit, and stellar activity.
}
{This scenario offers a viable pathway for the formation of volatile-rich atmospheres in evolved low-mass planets. When accompanied by dynamical signatures such as eccentric orbits, these chemical anomalies may trace past planetary disruption. This framework may help us interpret the atmospheric and orbital properties of systems such as TOI-421b, a warm sub-Neptune with a H/He-rich envelope and moderate eccentricity, and WASP-107b, a low-density Neptune-mass planet showing ongoing He I escape, shedding new light on the late-stage evolution of planetary systems.
}


\titlerunning{Volatile Enrichment from Tidal Disruption}

\authorrunning{Sucerquia et al.}
\maketitle
%

\section{Introduction}
\label{sec:intro}

Multiple lines of evidence suggest that close-in giant planets may eventually undergo tidal disruption. The ultra-short-period planets TOI-6255 b \citep{Dai2024} and BD+05 4868 Ab \citep{Hon2025}, both near their Roche limit -- the minimum distance at which a planet avoids tidal disruption by its host star -- may currently exemplify this process, offering rare real-time glimpses of planetary destruction. Similarly, chemical anomalies in the wide binary HD 240430 and HD 240429 have been attributed to planetary accretion by one star \citep{Oh2018}, while the transient event ZTF SLRN-2020 shows signatures consistent with the engulfment of a Jupiter-like planet by a main-sequence star \citep{Lau2025}. Such events can leave lasting compositional and dynamical imprints on the host star and surviving planets.

In the final stages of orbital decay \citep[e.g.,][]{Jackson2016, Spina2024, Alvarado-Montes2025}, a close-in giant planet may overflow its Roche lobe, releasing gas and dust into the inner system. While dust is expected to spiral inwards under Poynting–Robertson drag, the gas, primarily composed of hydrogen and helium, can be redistributed via viscous diffusion along the orbital plane \citep{Metzger2012}. This process drives mass inwards and angular momentum outwards, allowing a fraction of the volatile-rich material to migrate to wider orbits. There, it may interact with and be captured by surviving planets, potentially forming transient circumplanetary structures or modifying their atmospheric composition.

\begin{figure*}
  \centering
  \hfill
  \begin{minipage}[t]{0.33\textwidth}
    \vspace{0pt}
    \includegraphics[width=0.945\textwidth]{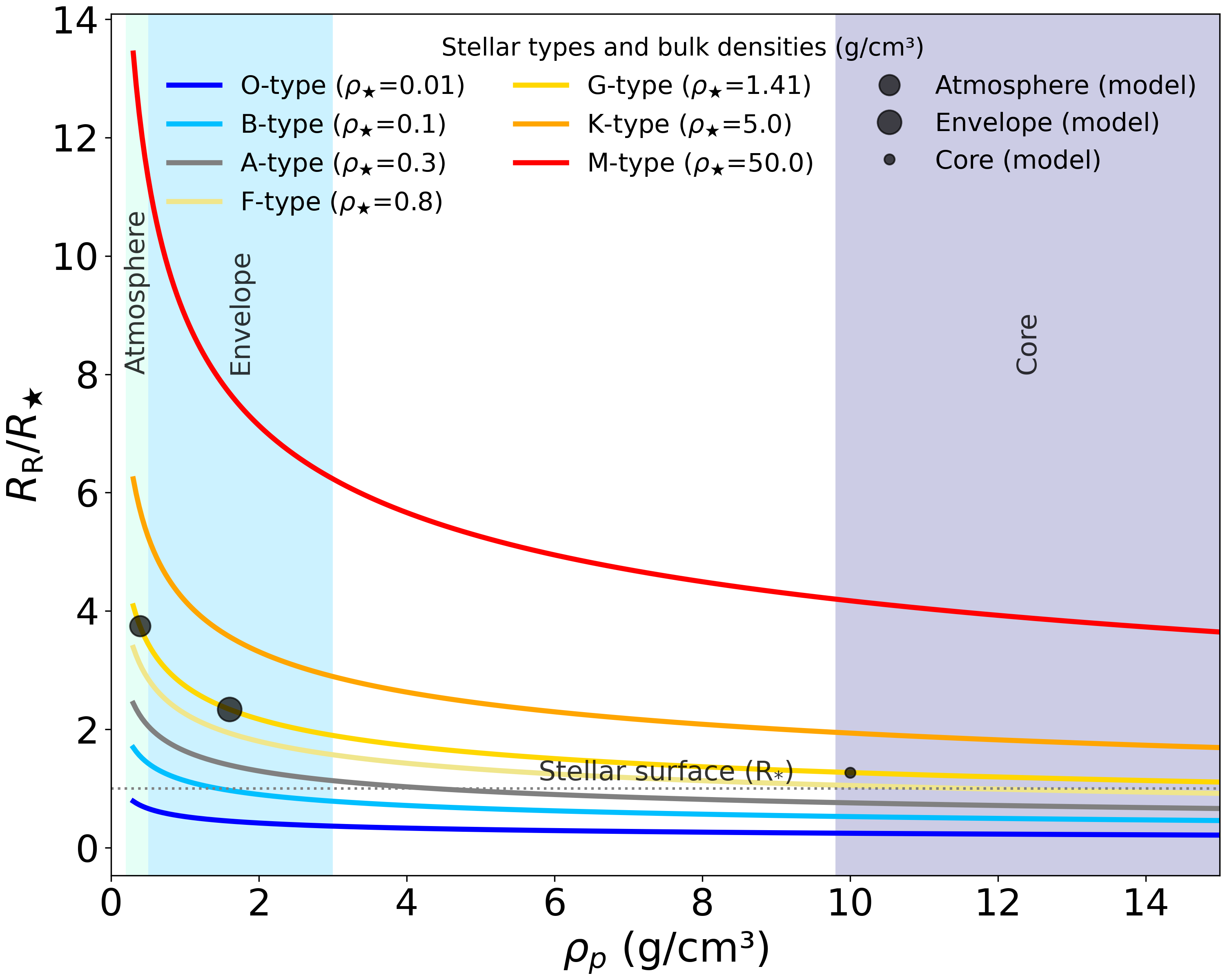}
  \end{minipage}%
  \hfill
  \begin{minipage}[t]{0.33\textwidth}
    \vspace{0pt}
    \includegraphics[width=0.945\textwidth]{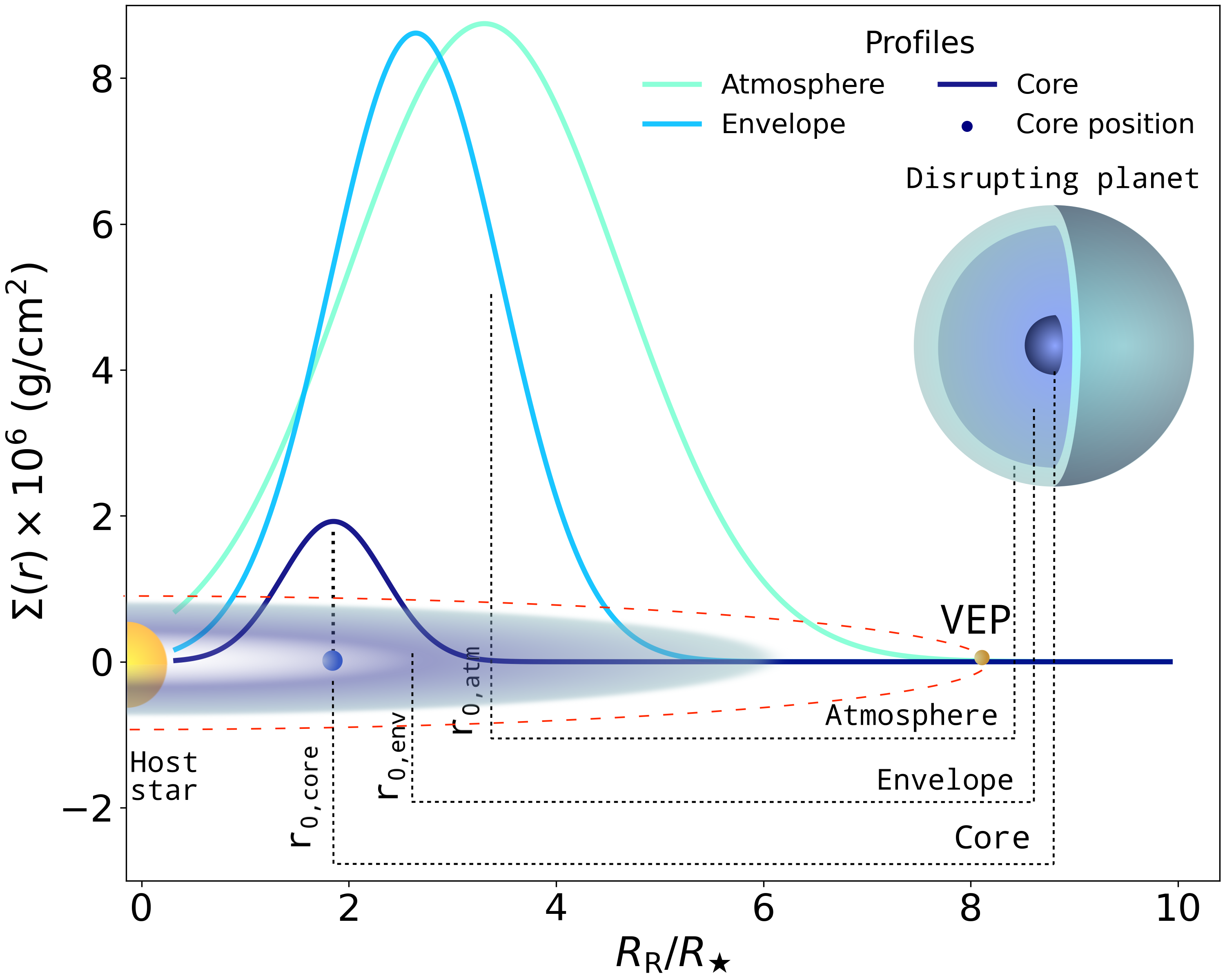}
  \end{minipage}%
  \hfill
  \begin{minipage}[t]{0.33\textwidth}
    \vspace{0pt}
    \caption{
      {(a)} Roche limit, $ R_{\rm R}  $, normalized by the stellar radius, $ R_\star  $ (see Appendix~\ref{app:ring}),
  $ R_\bigstar  $, 
      as a function of planetary bulk density, $ \rho_{\rm p}  $, for main-sequence stars of 
      different spectral types (O–M). The coloured bands represent the density ranges of a 
      three-layer Jupiter-like planet model, with black circles marking the Roche limit for 
      each layer used in this work. 
      {(b)} Surface density profiles, $ \Sigma(r)  $, of the disrupted atmosphere (green), 
      envelope (cyan), and core (dark blue), scaled to  $ R_\bigstar  $; the core’s location in the simulations is highlighted in dark blue.
    }
    \label{fig:diskmorph}
  \end{minipage}%
\end{figure*}

Several known planetary systems -- including WASP-47 \citep{Becker2015}, TOI-1130 \citep{Huang2020}, and TOI-2109 \citep{Harre2024} -- harbour both a close-in giant planet and a nearby low-mass planetary companion at wider orbital separations, indicating that inward migration and decay do not necessarily destabilise the planetary architecture \citep{Wu2023}. Notably, HAT-P-11\,b, a Neptune-mass planet on a moderately eccentric orbit, exhibits a low-density, hydrogen-rich atmosphere with confirmed He\,I absorption \citep[e.g.,][]{Allart2018}
, challenging models of an irradiated low-mass planet. While such signatures may arise from intrinsic processes, they could also reflect past mass transfer from a disrupted gas giant, supporting the idea that planetary break-up leaves lasting chemical imprints on surviving companions.

In this work, we investigate whether the disrupted envelope of a close-in giant planet can be accreted by a nearby low-mass planet, a VEP (volatile-enriched planet). Using 2D hydrodynamical simulations, supported by numerical $N$-body integrations and analytical models, we examine debris transport, gravitational capture conditions, and system stability (Sect.~\ref{sec:methodology}). We estimate the persistence of the resulting secondary atmospheres under photoevaporative escape and assess the detectability of their volatile content. By outlining a plausible evolutionary pathway (Sect.~\ref{sec:results}), we aim to clarify the origin of volatile-rich, inflated planets and identify spectroscopic and dynamical signatures of planetary disruption (Sect.~\ref{sec:discussion}).
\vspace{-0.48cm}
\section{Methodology}
\label{sec:methodology}
\paragraph{Gas ring and planet set-up:} We assumed that the tidal disruption of a close-in giant planet produces a circularised debris ring around the host star. The ring forms from the planet’s atmosphere and envelope, which are stripped upon exceeding their Roche limits -- $r_{0,\mathrm{atm}}$ and $r_{0,\mathrm{env}}$, respectively. $r_{0,\mathrm{i}}$ is the distance beyond which tidal forces overcome self-gravity. The denser core remains bound near  $r_{0,\mathrm{core}} $. The ring was modelled using a three-layer internal structure \citep{Miguel2022}, with the atmosphere and envelope as Gaussian surface density profiles centred at their disruption radii, and the core as a point mass (see Fig.\ref{fig:diskmorph} for a summary; Appendix \ref{app:ring} and Tables~\ref{tab:fiducial_params}–\ref{tab:sim_params} for further details).

We ran 2D hydrodynamical simulations with \texttt{FARGO3D} \citep{Benitez-Llambay+2016} on a logarithmically spaced radial grid and azimuthal equally spaced sectors over $2 \pi$, adopting a locally isothermal equation of state and a constant kinematic viscosity of $\nu = 7 \times 10^{13}\,\mathrm{cm^2\,s^{-1}}$. To study volatile capture, we placed a VEP candidate (1 or 10~$M_\oplus$) at $r = 0.04$ or $0.07$~au (equivalent to $\sim\!8.6$ and $15\,R_\star$, respectively, assuming $R_\star = 0.00465$~au). We hereafter refer to these as the `closer' and `farther' VEPs. We defined an effective capture radius of $ R_{\rm cap} = 0.2\,R_{\rm H}  $, an empirical scale derived from energy balance considerations of the envelope \citep{Montesinos2025}, where  $ R_{\rm H}  $ is the Hill radius of the VEPs. Gas entering this region was considered to be accreted, and the instantaneous envelope mass was computed as
\begin{equation}\label{eq:env_mass} 
    M_{\rm env}(t) = \int_{r < R_{\rm cap}} \Sigma(r, \phi, t)\, r\, dr\, d\phi. 
\end{equation}
Further details on the initial gas ring configuration, mass normalisation, and numerical set-up are provided in Appendix~\ref{app:ring}.
\vspace{-0.7cm}
\paragraph{Envelope contraction and gas accretion:} Gas accretion is regulated by outer envelope contraction. To estimate the effective accretion rate onto the outer planet under thermal contraction, we assumed that the captured mass, $M_{p}(t)$, evolves following a Kelvin-Helmholtz cooling prescription:
\begin{equation}\label{eq:mass_env}
    \dot{M}_\mathrm{p}(t) = M_\mathrm{env}(t) / t_\mathrm{KH},
\end{equation}
where $t_{KH}$ is the Kelvin-Helmholtz contraction timescale, defined as the characteristic time over which the planet’s envelope radiates away its gravitational binding energy and contracts. 
Following \citet{Ikoma+2000}, we approximated $t_{KH}$ as
\begin{equation}\label{eq:tau_KH}
    t_{KH} \sim 10^8 \left( \frac{M_{\rm core}}{M_{\oplus}} \right)^{1-1/q'} \left( \frac{\kappa_R}{1 \mathrm{cm}^2 \mathrm{g}^{-1}} \right)^{s/q'} \mathrm{yr}, 
\end{equation}
where $M_{\rm core}$ is the core mass of the outer planet and the numerical exponents are set to $1 - 1/q' \simeq -2.5$ and $s/q' \simeq 1$. For the opacity we used the Rosseland mean opacity, $\kappa_R$, derived from \cite{Semenov2003}, relevant to the physical conditions of protoplanetary discs. 
\vspace{-0.4cm}
\paragraph{Pre- and post-disruption dynamical architecture:} Understanding the stability of VEPs is crucial both before and after the disruption of a close-in gas giant. While hot Jupiters were once thought to be dynamically isolated, recent discoveries of nearby companions -- detected via transit timing variations (TTVs), for example -- challenge this view \citep[e.g.,][]{Becker2015}. Prior to disruption, resonant coupling can drive collective inward migration \citep[e.g.,][]{Cerioni_2023}, shaping the initial system architecture. These resonances not only stabilise planetary orbits but also open pathways for material transfer, potentially enabling VEPs to accrete gas from the inner planet’s envelope. After disruption, however, the shifting stellar mass and remnant core perturbations may trigger dynamical instabilities, exciting eccentricities, inducing resonant captures, or even leading to ejection or collision events.

To explore these effects, we performed a suite of $N$-body simulations using \texttt{REBOUND} \citep{Rein2012}, computing the mean exponential growth factor of nearby orbits (MEGNO) as a chaos indicator. We examined two regimes: close-in orbits near the remnant’s location -- dominated by mean motion resonances (MMRs) -- and more distant regions where outer planets initially reside. We also modelled the gradual fading of a hot Jupiter, transferring 95\% of its mass to the star over 20 years, to study the dynamical response of exterior Earth-like planets.
\vspace{-0.4cm}
\section{Results and analysis} 
\label{sec:results}

\subsection{Disc evolution and volatile capture}

\begin{figure*}
  \centering
    \includegraphics[width=0.32\textwidth]{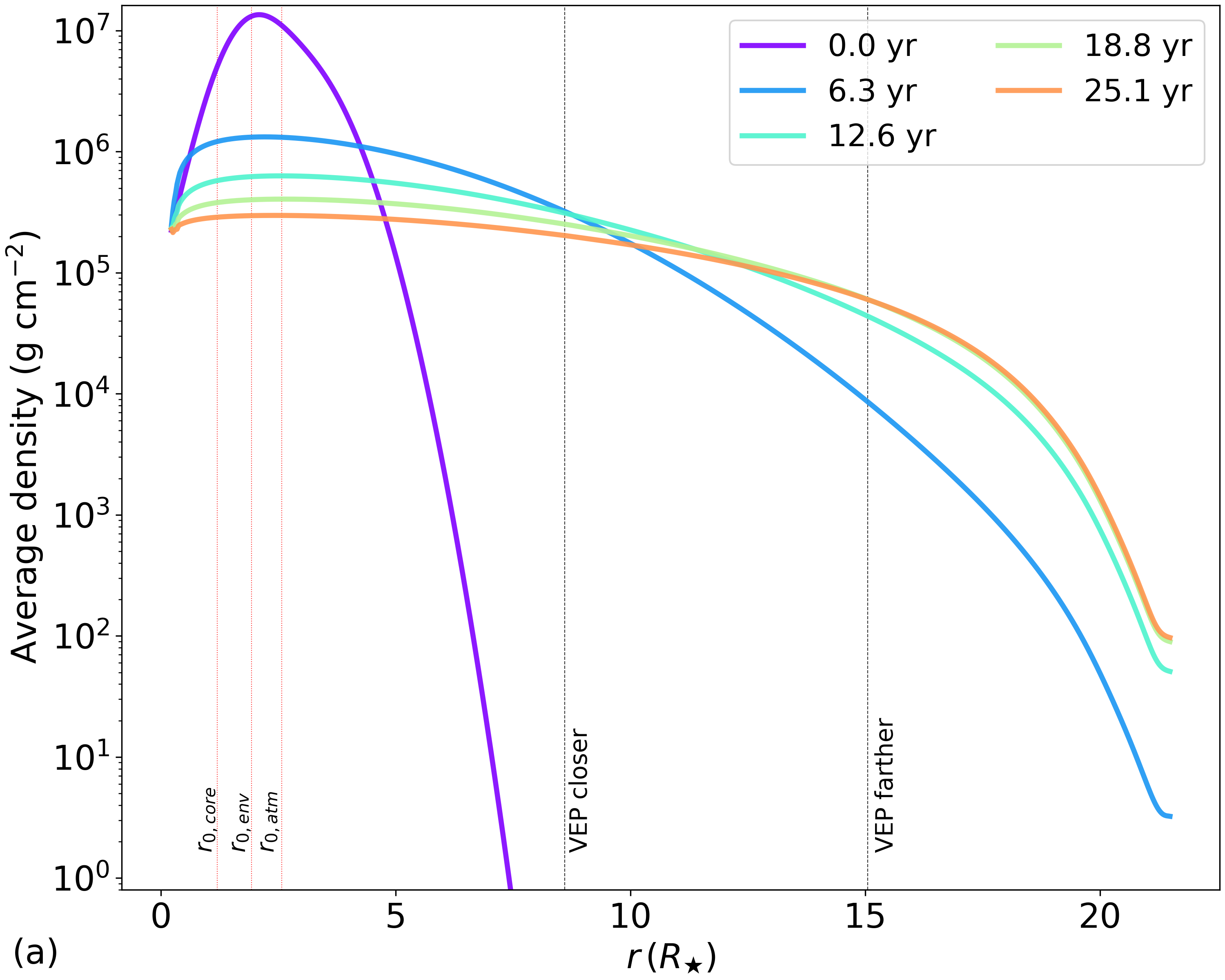}
  \includegraphics[width=0.32\textwidth]{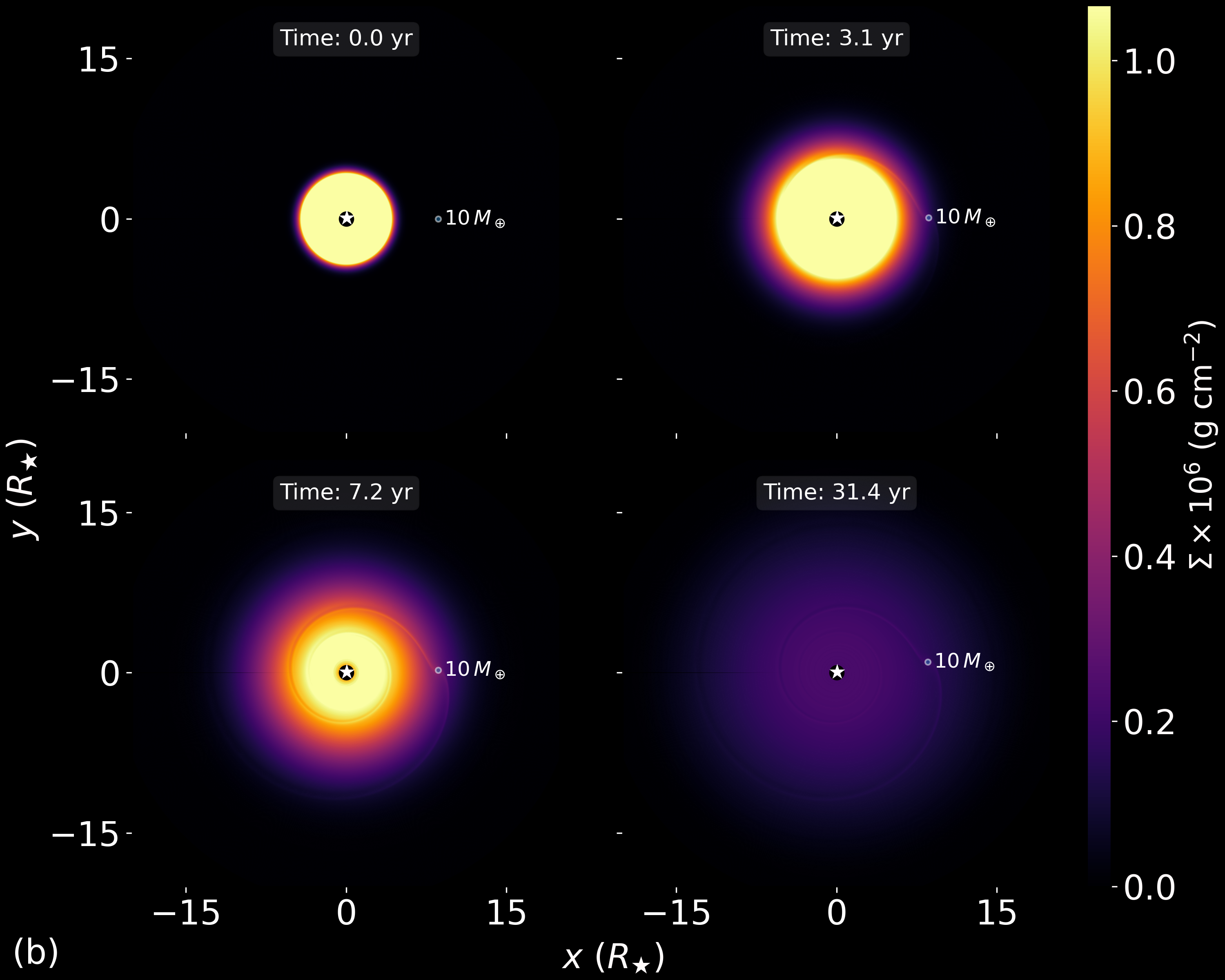}
   \includegraphics[width=0.32\textwidth]{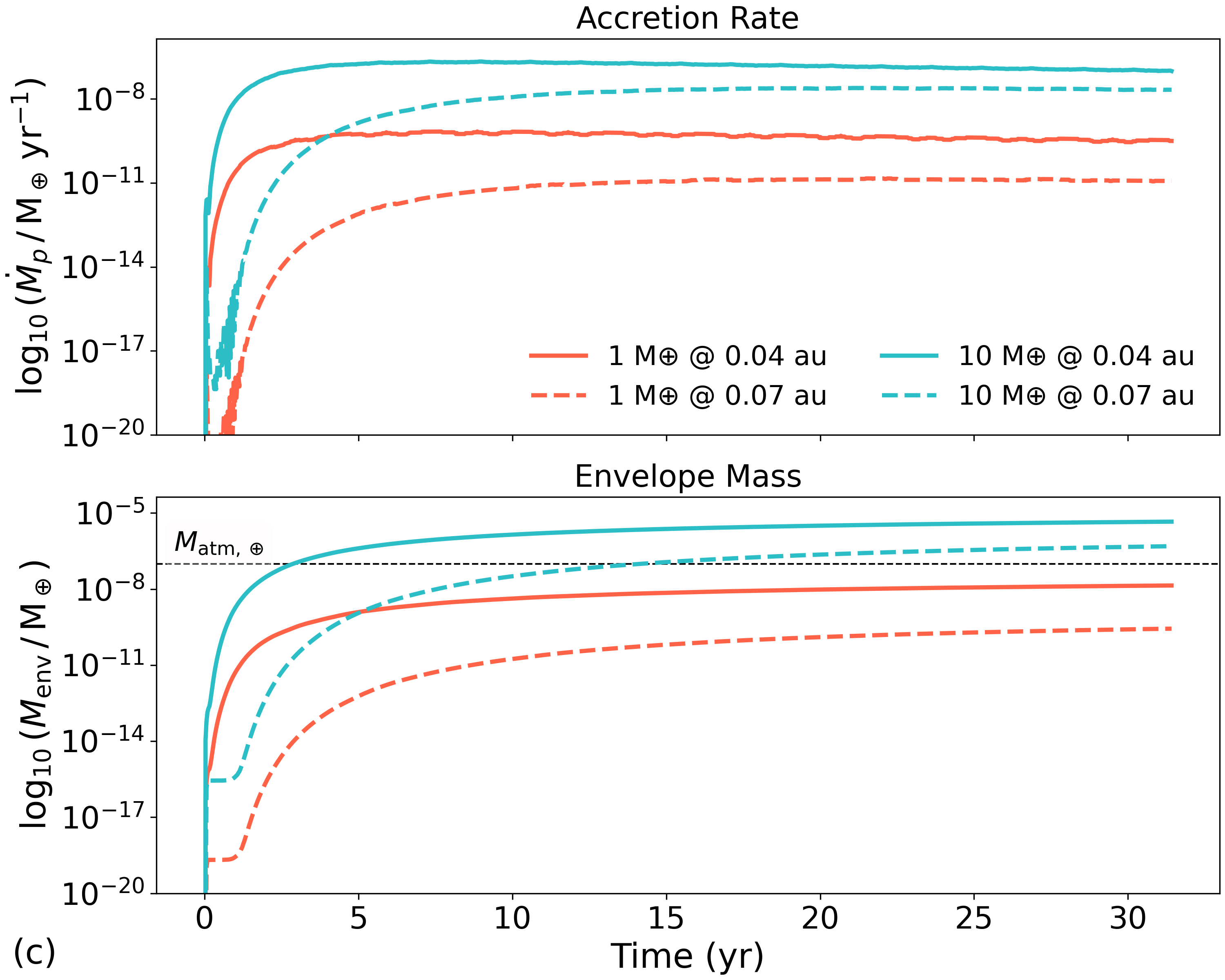}
  \caption{Time evolution of the gas disc and planetary envelopes after the tidal disruption of a Jupiter-like planet. (a) Azimuthally averaged surface density at four times; dashed lines mark the {closer} and {farther} VEPs. (b) Disc surface density evolution for the  $10\,M_\oplus $ case; circles indicate the {closer} VEP. (c): Accretion rates (top) and envelope masses (bottom) for VEPs at both orbits. $M_{\mathrm{atm},\oplus}$ indicates the atmospheric mass of Earth.}
  \label{fig:density_fields_models}
\end{figure*}

Hydrodynamical simulations show that the debris ring produced by the tidal disruption of a Jupiter-like planet rapidly spreads outwards due to viscosity over a few years. Figure~\ref{fig:density_fields_models}.a displays the time evolution of the azimuthally averaged surface density profile for the disrupted gas. Initially concentrated near the disruption radius ($\sim\!0.065$~au), the material diffuses outwards efficiently. Within the first $\sim\!3$ years, gas reaches the location of the closer VEP, enabling envelope accumulation, as is seen in the accretion rate peak of Fig.~\ref{fig:density_fields_models}.b. At later times, the surface density at this location steadily declines, smoothing out the local density enhancement, and the accretion rate onto the planet becomes constant, marking the onset of a quasi-steady viscous regime.

The 2D density field evolution for the simulation with a planetary core mass of 10~$M_\oplus$ is shown in Fig.~\ref{fig:density_fields_models}.b, illustrating the gas distribution following the tidal disruption event. In this case, the higher planetary mass induces transient spiral arms as the gas disperses ($t \sim 31.4$ yr), highlighting the dynamical influence of the massive companion on the surrounding flow.
The gas entering the vicinity of the outer planet defines the available envelope mass. The subsequent evolution of this envelope is governed by its ability to radiate away its internal energy and contract (Eq.~\ref{eq:mass_env}). The efficiency of this contraction depends on the envelope opacity, described by the Kelvin-Helmholtz timescale (Eq.~\ref{eq:tau_KH}).
The accretion rate and resulting envelope masses for the closer and farther VEPs are shown in Fig.~\ref{fig:density_fields_models}.c. More massive planets consistently exhibit a higher accretion efficiency, leading to more substantial envelopes, largely independent of their orbital distance. Planets located closer to the disruption zone tend to accumulate more material, but even those farther out retain a non-negligible fraction of gas. Quantitative values are provided in Table.~\ref{table:ppm}.

\begin{table}[]
\centering
\setlength{\tabcolsep}{4pt}
\renewcommand{\arraystretch}{0.5}
\small
\begin{tabular}{ccccc}
\toprule
$a_\mathrm{p}$ ($R_\star$) & Mass ($M_\oplus$) & Captured mass ($M_\oplus$) & $A_H$ (km) & $\Delta F$ (ppm) \\
\midrule
8.60  & 1  & $1.39\times10^{-8}$  & $1.01\times10^{4}$ & 292.99 \\
8.60  & 10 & $4.58\times10^{-6}$  & $1.21\times10^{3}$ &  35.06 \\
15.05 & 1  & $2.75\times10^{-10}$ & $6.28\times10^{3}$ & 182.08 \\
15.05 & 10 & $4.93\times10^{-7}$  & $8.33\times10^{2}$ &  24.15 \\
\bottomrule
\end{tabular}
\caption{Captured-envelope masses and transit depths for VEPs.}
\label{table:ppm}
\vspace{-0.2cm}
\end{table}
\vspace{-0.3cm}

\subsection{Transit spectroscopy of captured volatiles}
A key implication of our scenario is that hydrogen- and helium-rich gas released during the tidal disruption of a close-in giant planet can be gravitationally captured by a VEP, forming a transient secondary atmosphere. This type of volatile enrichment, unexpected in evolved terrestrial planets, can imprint detectable spectral features during planetary transits. To assess the observability of such atmospheres, we estimated the effective height of the captured gas envelope ($A_H$, derived for isothermal atmospheres under hydrostatic equilibrium) and the corresponding transit depth ($\Delta F$, e.g. \citealt{deWit2013}).
The atmospheric contribution to the transit depth is given by
\begin{equation} \label{eq:deltaf}
\Delta F \approx \frac{2\,R_p\,A_H}{R_*^2},
\end{equation}
where $R_p$ is the planetary radius, $R_*$ the stellar radius, and $A_H$ the effective height at which the atmosphere becomes optically thick in the tangential direction. The derivation of the effective atmospheric height and its dependence on atmospheric mass, opacity, and scale height is presented in Appendix~\ref{ap:detection}.

According to our model, the effective envelope height reaches $A_H \sim 6,000$–$10,000$ km for $1\,M_\oplus$ planets, and $\sim 800$–$1,200$ km for $10\,M_\oplus$ planets. This trend reflects the weaker surface gravity of low-mass planets, which allows the captured gas to remain more extended. The resulting transit depths range from approximately 24 to 293 ppm across the explored configurations. While modest, these signals fall within the sensitivity of current and upcoming high-precision instruments such as JWST and Ariel \citep[see review by][]{Tinetti2022}. Table~\ref{table:ppm} summarises these results, and the temporal evolution of $\Delta F$ during envelope capture is illustrated in Appendix~\ref{ap:detection}.

Our simulations show that low-mass planets ($1$–$10,M_\oplus$) can capture hydrogen-dominated gas from the disrupted giant, forming extended secondary envelopes. These atmospheres can reach effective heights of several thousand kilometres and produce transit depths of up to a few hundred parts per million. The extent of the envelope scales inversely with planetary mass, due to weaker surface gravity, and the signals remain within the detection thresholds of current and upcoming observatories.
\vspace{-0.4cm}
\subsection{Persistence of captured volatile atmospheres}\label{subsec:Persistence}
We estimated the long-term persistence of captured H/He envelopes using the mass-loss formalism given in Eq.~(12) of \citet{Alvarado-Montes2025}, which accounts for both XUV-driven escape and stellar wind drag. In mature systems in which the host star has exited its XUV-saturated phase, the survival of a captured envelope is greatly extended. At the closer and farther VEP orbits, where $F_{\rm XUV} \sim 1$--10 erg\,s$^{-1}$\,cm$^{-2}$, the mass-loss rates drop to $\sim3\times10^{-14}$--$3\times10^{-15}\,M_\oplus\,\mathrm{yr^{-1}}$. Given envelope masses of $3\times10^{-10}$--$3\times10^{-6}\,M_\oplus$, the atmospheric persistence timescales reach $10^6$--$10^8$ years, allowing such transient volatile layers to survive well beyond typical disc lifetimes. Therefore, they may be observable even millions of years after the original disruption and capture event.
\vspace{-0.4cm}
\subsection{Disc dispersal timescale}\label{subsec:DiscDispersal}
The gas released during the disruption spreads outwards under viscous evolution, and the characteristic dispersal timescale is given by $t_\nu \sim r^2 / \nu$. At the closer VEP's location and with $\nu = 7 \times 10^{13} \mathrm{cm^2\,s^{-1}}$, we obtain $t_\nu \approx 163 \mathrm{yr}$. This indicates that the disc is rapidly depleted. However, by the time this occurs, the planet has already acquired an envelope of $\sim 3 \times 10^{-11}$–$3 \times 10^{-6}\,M_\oplus$ (see Fig.~\ref{fig:density_fields_models}.c). As was discussed earlier, these envelopes are resilient to stellar irradiation: their evaporation timescales, $\tau_{\rm esc}$, span $10^6$–$10^{8}$ yr, ensuring that the captured gas survives long after the disc has dissipated.

\vspace{-0.4cm}
\subsection{Orbital evolution}
The $N$-body simulations indicate that the system architecture is strongly influenced by mean-motion resonances (MMRs) with the inner remnant (Appendix~\ref{app:dynamics}). Two-dimensional MEGNO maps (Fig.~\ref{fig:MENGOevol}) identify bands of dynamical chaos near low-order resonances (2:1, 3:2, 5:3), particularly within $\sim0.015$ au, where eccentricities increase from initially zero to $\sim0.07$ (Fig.~\ref{fig:plan2dinevol}). Such excitation can enhance gas loss, alter accretion processes, or induce engulfment of the remnant core.

Beyond $\sim0.015$ au, the system remains dynamically cold, supporting the conservative yet physically motivated selection of VEP locations. In these regions, gas interactions are relatively mild, enabling passive volatile capture without triggering significant migration or dynamical instability.

During the tidal disruption phase, the fading planet simulations reveal modest eccentricity growth and slight inward migration of both the outer planets and the remnant core, driven by stellar mass increase. Although resonant coupling and mass accretion can sustain eccentricity excitation, most configurations at the closer and farther VEP orbits retain dynamical stability, confirming their viability as potential VEPs. Further details and supplementary figures are provided in Appendix~\ref{app:dynamics}.

\vspace{-0.5cm}
\section{Discussion and conclusions}
\label{sec:discussion}

We have shown that tidal disruption of close-in gas giants can deliver volatile-rich gas to a VEP, forming transient H/He envelopes with distinct spectral signatures, as are revealed by analytical and hydrodynamical models. Such envelopes, though tenuous, may be detectable through transmission spectroscopy, especially if they are hydrogen-rich and extended.

Recent high-resolution models of hot Jupiter atmospheres \citep{Kafle2025} confirm the sensitivity of spectral features to minor atmospheric components, supporting the feasibility of detecting low-mass volatile layers. This mechanism could help explain the origin of some puffy planets, whose inflated envelopes may reflect recent volatile capture or inefficient contraction -- though alternative pathways remain possible.

In this scenario, while the disrupted core is engulfed by the host star, the VEP may re-accrete some of the expelled gas. In G-type stars with shallow convective envelopes, such ingestion could measurably raise the stellar [Fe/H] \citep{Oh2018}, whereas the VEP acquires a low-metallicity H/He layer. This compositional divergence, possibly accompanied by moderate eccentricity induced by dynamical interactions -- as is also suggested by our N-body simulations -- may trace the imprint of past disruption events and suggest a chemical decoupling between stars and their planets (see Appendix~\ref{app:dynamics} for orbital details).

Several observed systems align well with the predictions of our scenario. WASP-107b \citep{Krishnamurthy2025} and HAT-P-11b \citep{Allart2018} are warm, low-density Neptunes with extended H/He-rich atmospheres and eccentric orbits ($e \gtrsim 0.1$), hosted by stars with super-solar metallicities. This combination is consistent with a past tidal disruption event in which the remnant core was accreted by the star and a volatile envelope was re-accreted by a VEP. TOI-421b, in contrast, shows a similar planetary configuration -- a hydrogen-rich atmosphere and $e \simeq 0.13$ -- but its host star presents near-solar metallicity \citep{Davenport2025}. This discrepancy may reflect either the absence of ingestion or rapid dilution in the shallow convective envelope of its G9/K0V host \citep{Spina2024}. Although not conclusive, these examples highlight that disruption-driven volatile re-accretion remains a viable origin pathway for some inflated sub-Neptunes.

Additionally, the morphology of re-accreted gas -- potentially extended, asymmetric, or optically thick -- may produce anomalous transit signatures that mimic circumplanetary structures like rings or discs, leading to misclassification or overestimated planetary radii. Recent JWST/NIRSpec transmission studies of Kepler-51d \citep{Libby-Roberts2025} show how low-density planets can present degenerate signals, where inflated H/He envelopes, high-altitude hazes, or tilted rings yield similar observational features \citep{Sucerquia2017}. These cases underscore the need for careful modelling when diagnosing volatile-rich atmospheres formed via planetary disruption.

Alternative sources of volatiles, such as cometary impacts or internal degassing, can also contribute to secondary atmospheres in low-mass planets. However, these mechanisms are unlikely to yield the H/He-rich compositions expected from tidal disruption. Targeting spectral features such as He I 10830~\AA, H$\alpha$, or CH$_4$ in planets orbiting inactive stars may help to distinguish these origins, using high-sensitivity instruments such as JWST, ELTs, or ESPRESSO. Our results show that these captured envelopes can survive XUV-driven escape without requiring magnetic shielding. Still, accounting for such effects could further extend their persistence, chiefly in more active systems \citep{Zuluaga2013}.

In summary, our results suggest that some warm Neptunes with H/He-rich atmospheres, eccentric orbits, and metal-enriched host stars may be relics of past tidal disruption events. While hydrogen is ubiquitous and multiple pathways may produce similar signatures, the joint presence of dynamical and compositional anomalies offers a promising diagnostic. Our scenario thus complements existing formation models and provides a new lens to interpret outliers in planetary architecture and composition.

\begin{acknowledgements}
   This project was supported by the European Research Council (ERC) under the European Union Horizon Europe research and innovation program (grant agreement No. 101042275, project Stellar-MADE). M.M. acknowledges financial support from FONDECYT Regular 1241818.
\end{acknowledgements}
    
\vspace{-0.5cm}
\bibliographystyle{aa}
\vspace{-0.4cm}
\bibliography{biblio}

\appendix
\section{Gas ring parameters and numerical set-up}\label{app:ring}

To determine the gas distribution in the debris ring, we compute the local Roche radius $r_{0,\mathrm{i}}$ for each planetary layer (atmosphere, envelope, and core) as\begin{equation}\label{eq:roche_radius_app}
    r_{0,\mathrm{i}} = 2.44\, R_\bigstar \left( \frac{\rho_\bigstar}{\rho_{\mathrm{p},i}} \right)^{1/3},
\end{equation}
where  $ \rho_\bigstar  $ is the mean stellar density, and  $ \rho_{\mathrm{p},i}  $ is the bulk density of the corresponding planetary layer.

The Roche radius defines the critical distance from the star at which a planetary component becomes gravitationally unbound due to tidal forces. Since the Roche radius increases with decreasing density, each layer is stripped at a different distance: the atmosphere, being the least dense, is lost first at  $r_{0,\mathrm{atm}} $; the envelope follows at  $r_{0,\mathrm{env}} $; and the denser core remains intact until it approaches  $r_{0,\mathrm{core}} $. 

Material exceeding its Roche radius is redistributed into a circumstellar ring, represented by Gaussian surface density profiles,
\begin{equation}
    \Sigma(r) = \sum_{i} \Sigma_{0,i} \exp\left( -\frac{(r - r_{0,i})^2}{2 \sigma_i^2} \right),
\end{equation}
where  $ i \in \{\mathrm{atm}, \mathrm{env}\}  $,  $ r_{0,i}  $ is the disruption radius,  $ \sigma_i  $ the radial extent, and  $ \Sigma_{0,i}  $ is normalised such that
\begin{equation}
   M_i = 2\pi \int_{r_{0,i} - 3\sigma_i}^{r_{0,i} + 3\sigma_i} \Sigma_i(r)\, r\, dr. 
\end{equation}The fiducial parameters adopted for the disrupted Jupiter-like planet are summarised in Table~\ref{tab:fiducial_params}.

\begin{table}[h]
\centering
\caption{Fiducial parameters for the disrupted Jupiter-like planet.}
\label{tab:fiducial_params}
\small
\begin{tabular}{lcccc}
\hline\hline
Component & $M_i$ ($M_{\rm Jup}$) & $r_{0,i}$ (au) & $\sigma_i$ (au) & $\Sigma_{0,i}$ (g\,cm$^{-2}$) \\
\hline
Atmosphere & 0.65 & 0.012 & 0.004 & $7.29 \times 10^6$ \\
Envelope     & 0.32 & 0.009 & 0.0025 & $7.66 \times 10^6$ \\
Core       & 0.02 & 0.0056 & -- & -- \\
\hline
\end{tabular}
\end{table}

Simulations were performed with a logarithmically spaced radial grid of  $ N_r = 720  $ and azimuthal grid of  $ N_\phi = 1024  $, covering  $ r_{\rm in} = 0.001 \,\mathrm{au}  $ to  $ r_{\rm out} = 0.1\,\mathrm{au}  $. A locally isothermal equation of state is adopted, with a constant kinematic viscosity of  $ \nu = 7 \times 10^{13}\,\mathrm{cm^2\,s^{-1}}  $. The fiducial parameters used in the simulations are summarised in Table~\ref{tab:sim_params}.

\begin{table}
  \centering
  \scriptsize
    \caption{Fiducial parameters used in simulations.}
      \label{tab:sim_params}
  \begin{tabular}{lll}
    \hline\hline
    \multicolumn{3}{c}{{Stellar (Solar-type)}}\\
    Mass & $M_\star$ & $1.0\,M_\odot$ \\
    Radius & $R_\star$ & $1.0\,R_\odot$ (0.00465\,au) \\
    \hline
\multicolumn{3}{c}{{Disrupted Jupiter-like planet}}\\
Total mass & $M_{\rm p}$ & $0.99\,M_{\rm Jup}$ \\
\makecell[l]{Layers \\ (atmosphere / mantle / core)} & $M_i$ & 0.65 / 0.32 / 0.02\,$M_{\rm Jup}$ \\
Disruption radii (Roche) & $r_{0,i}=R_{\rm Roche,i}$ & 0.012 / 0.009 / 0.0056\,au \\
\hline
    \multicolumn{3}{c}{{VEPs (volatile-enriched planets)}}\\
    Mass & $M_p$ & 1, 10\,$M_\oplus$ \\
    Semi-major axis & $a$ & 0.04, 0.07\,au \\
    Hill radius & $R_{\rm H}=a\,(M_p/3M_\star)^{1/3}$ & 
    \begin{tabular}{@{}l@{}}$4.0\times10^{-4}$\,au (1\,$M_\oplus$)\\
                         $1.5\times10^{-4}$\,au (10\,$M_\oplus$)\end{tabular}\\
    Capture radius & $R_{\rm cap}=0.2\,R_{\rm H}$ & 
    \begin{tabular}{@{}l@{}}$8.0\times10^{-5}$\,au (1\,$M_\oplus$)\\
                         $3.0\times10^{-5}$\,au (10\,$M_\oplus$)\end{tabular}\\
    \hline
    \multicolumn{3}{c}{{Disc and hydrodynamic simulation}}\\
    Kinematic viscosity & $\nu$ & $7\times10^{13}\,\mathrm{cm^2\,s^{-1}}$ \\
    Radial domain & $[r_{\rm in},r_{\rm out}]$ & [0.001,\,0.10]\,au \\
    Resolution & $(N_r,N_\phi)$ & (720,\,1024) \\
    Equation of state &  & locally isothermal \\
    \hline
    \multicolumn{3}{c}{{Kelvin--Helmholtz contraction}}\\
    Rosseland mean opacity & $\kappa_R$ & 1\,cm$^2$\,g$^{-1}$ \\
    Exponent $1-1/q'$ &  & $-2.5$ \\
    Exponent $s/q'$ &  & $1$ \\
    Time-scale & $t_{\rm KH}$ & 
    $\sim\,10^8\,(M_{\rm core}/M_\oplus)^{-2.5}\,\mathrm{yr}$ \\
    \hline
  \end{tabular}
\end{table}

\section{Dynamical survey results} \label{app:dynamics}

We performed a suite of $N$-body simulations using the \texttt{WHFAST} symplectic integrator, as implemented in the {\tt REBOUND} package \citep{Rein2012}, to investigate the dynamical response of outer low-mass planets to the disruption of a close-in gas giant. The system consists of a solar-mass star, a hot Jupiter initially at $a = 0.0065$\,au, and five secondary planets (1 and 10~$M_\oplus$) on circular orbits between 0.04 and 0.07\,au. Simulations use a fixed timestep of $\Delta t = 10^{-5}$\,yr, conserving energy and angular momentum over 10,000 years.
\begin{figure}
  \centering
  \includegraphics[width=0.9\columnwidth]{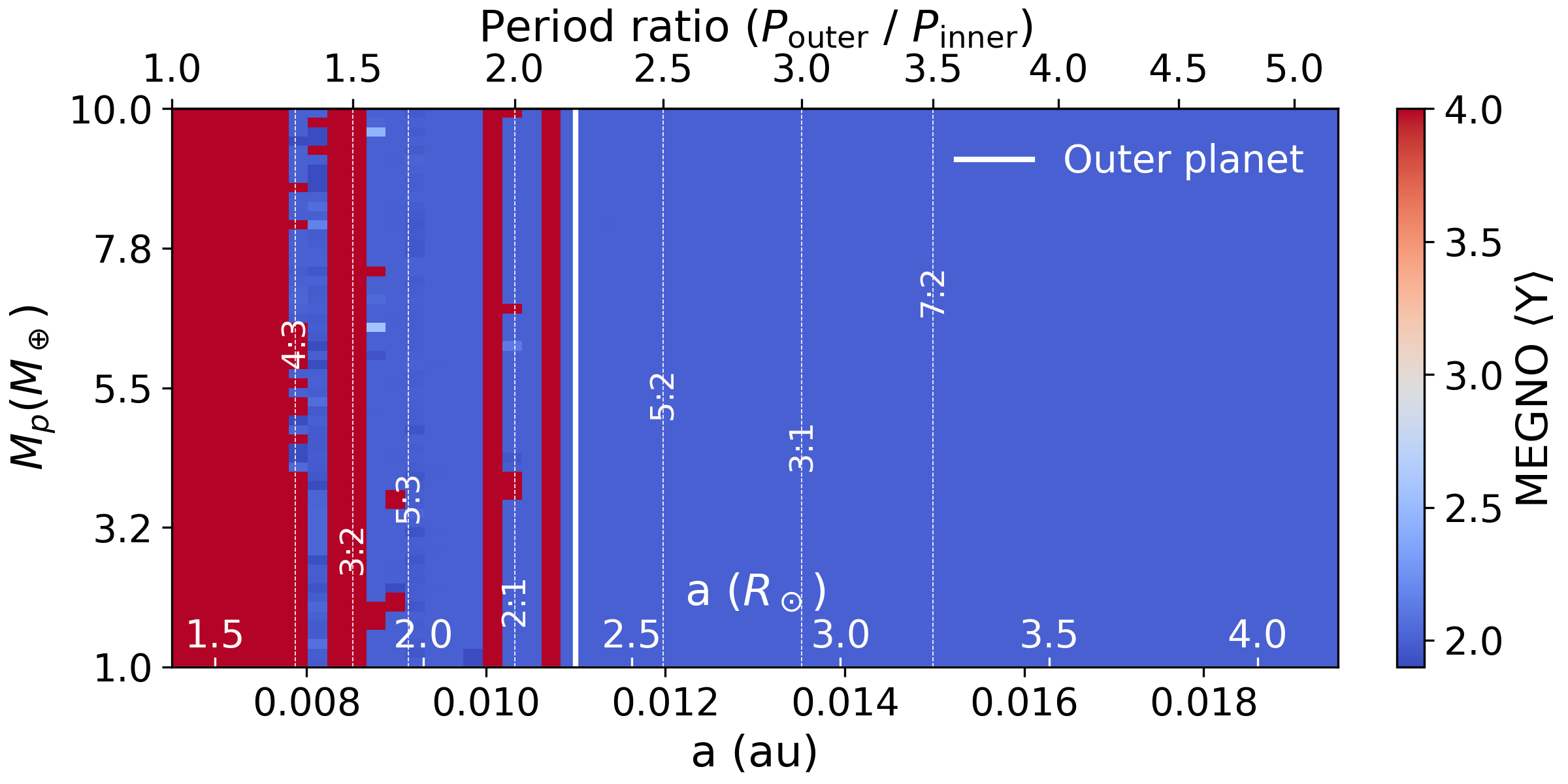}
  \includegraphics[width=0.9\columnwidth]{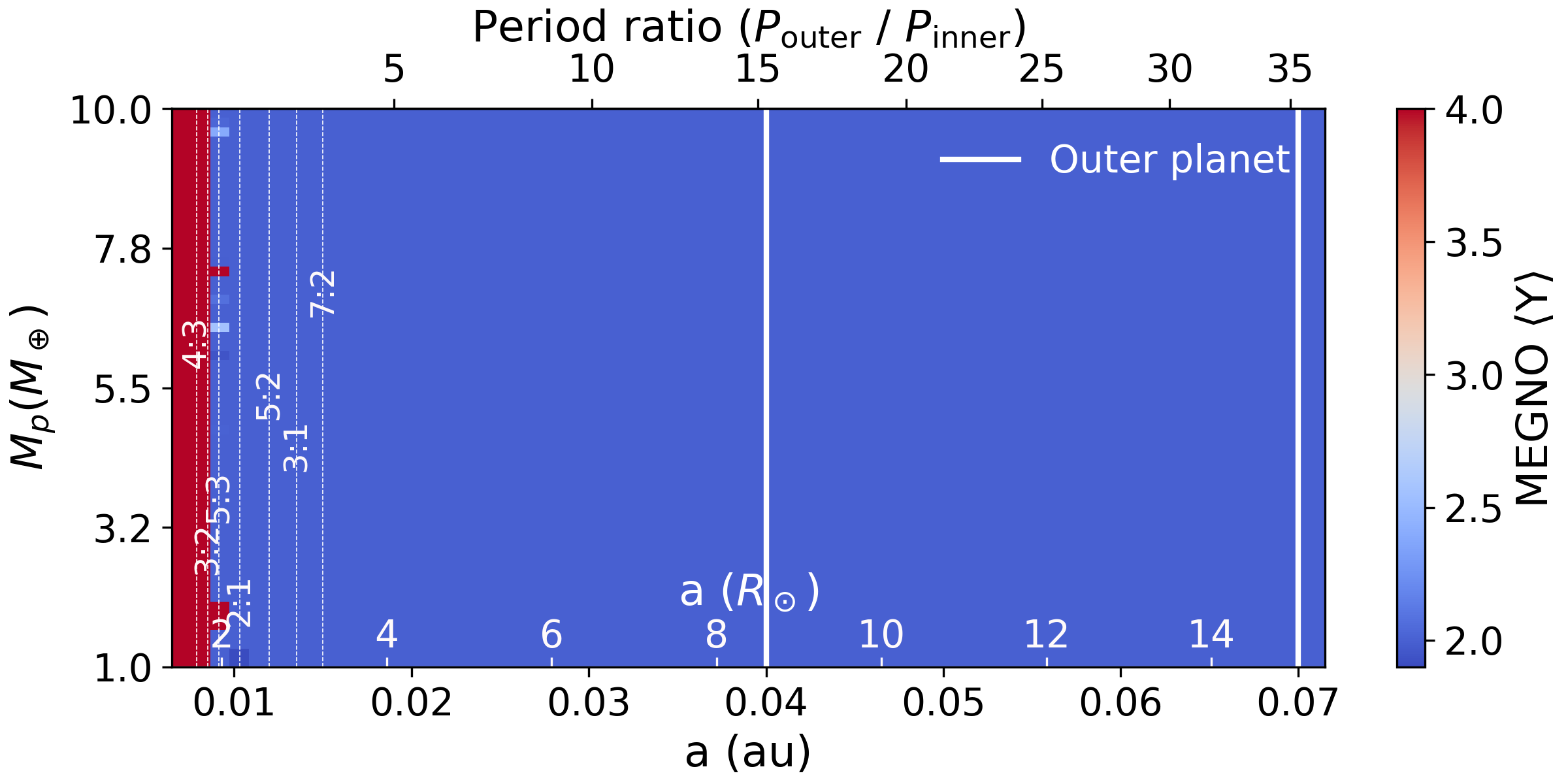}
  \caption{MEGNO maps showing dynamical structure as a function of outer planet semi-major axis and mass. Top: Closer regime ($<0.015$\,au), with strong chaotic zones near low-order resonances. Bottom: Farther regime ($>0.015$\,au), where most configurations remain regular.}
  \label{fig:MENGOevol}
\end{figure}

To map stable and chaotic zones, we first computed 2D MEGNO maps \citep{Cincotta2000} as a function of planetary mass and semi-major axis (Fig.~\ref{fig:MENGOevol}). The closer regime (top panel) exhibits strong chaotic behaviour near low-order MMRs (2:1, 3:2, 5:3), where eccentricity excitation is likely. In contrast, the farther regime (bottom panel), beyond  $\sim 0.015\,\mathrm{au} $, remains largely regular, supporting our choice of placing outer planets at 0.04–0.07\,au (closer and farther models) in the main simulations.

\begin{figure*}
 \label{fig:plan2dinevol}
  \centering
  \includegraphics[width=0.99\columnwidth]{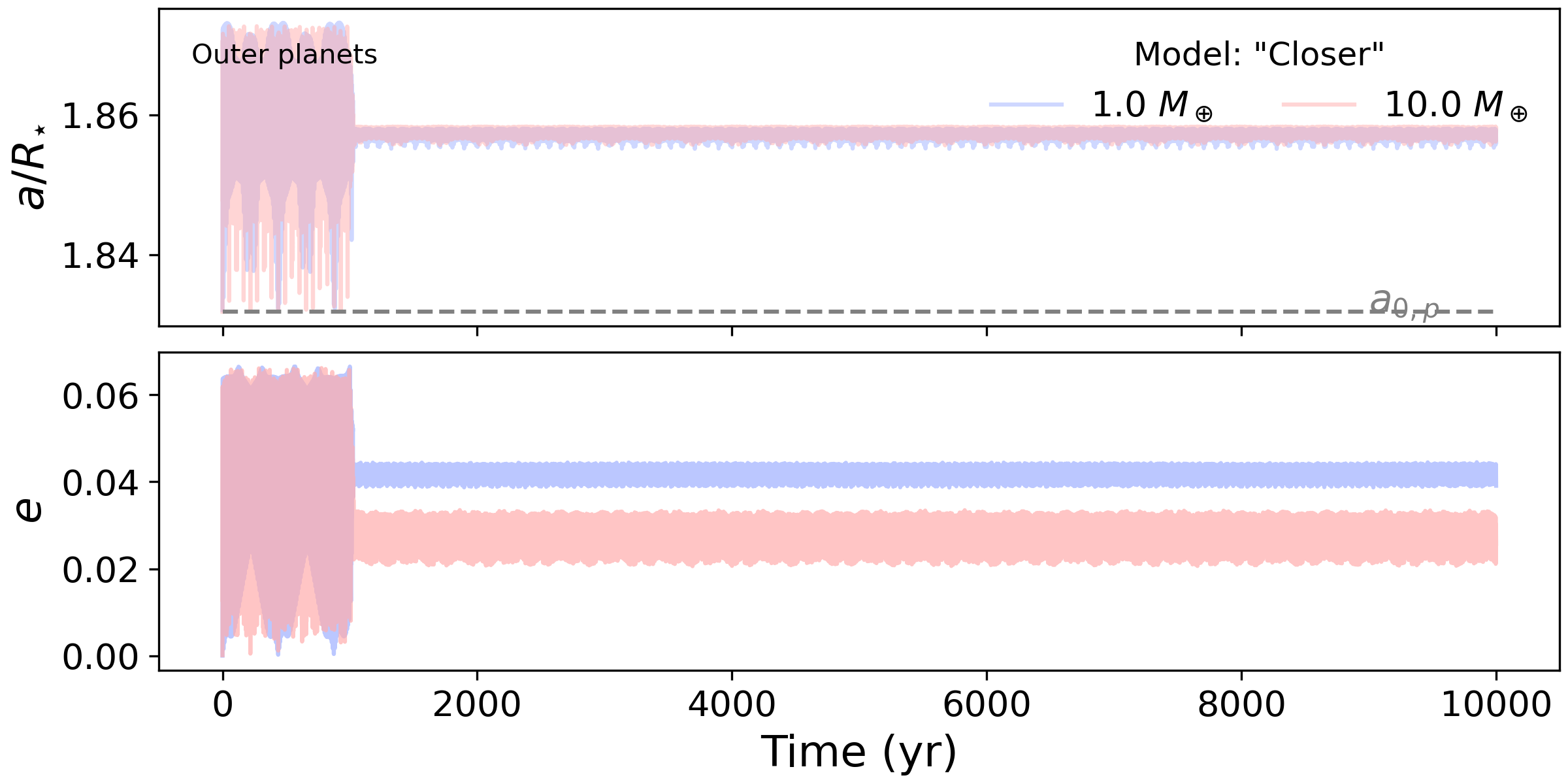}
  \includegraphics[width=0.99\columnwidth]{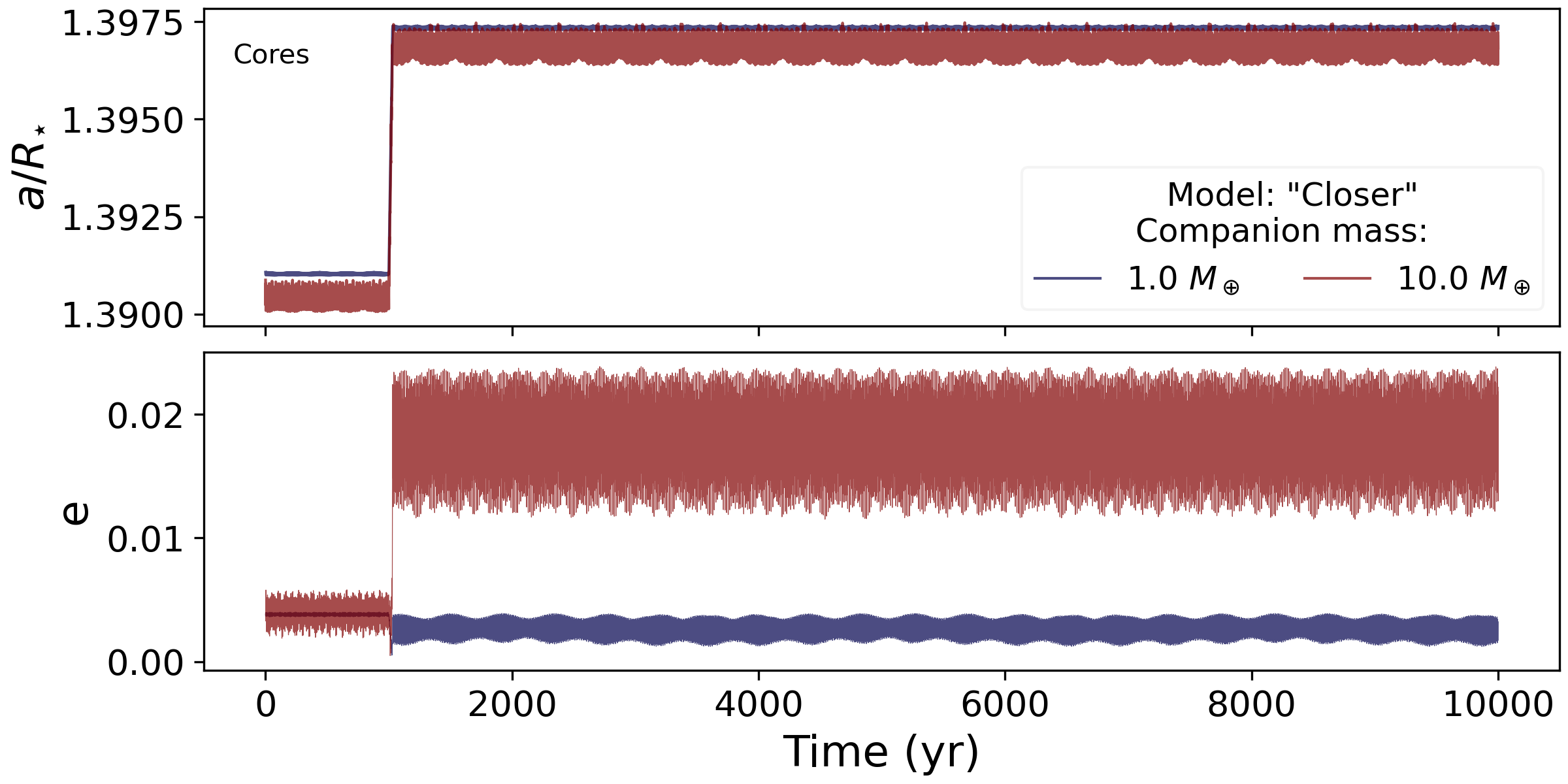}
  \includegraphics[width=0.99\columnwidth]{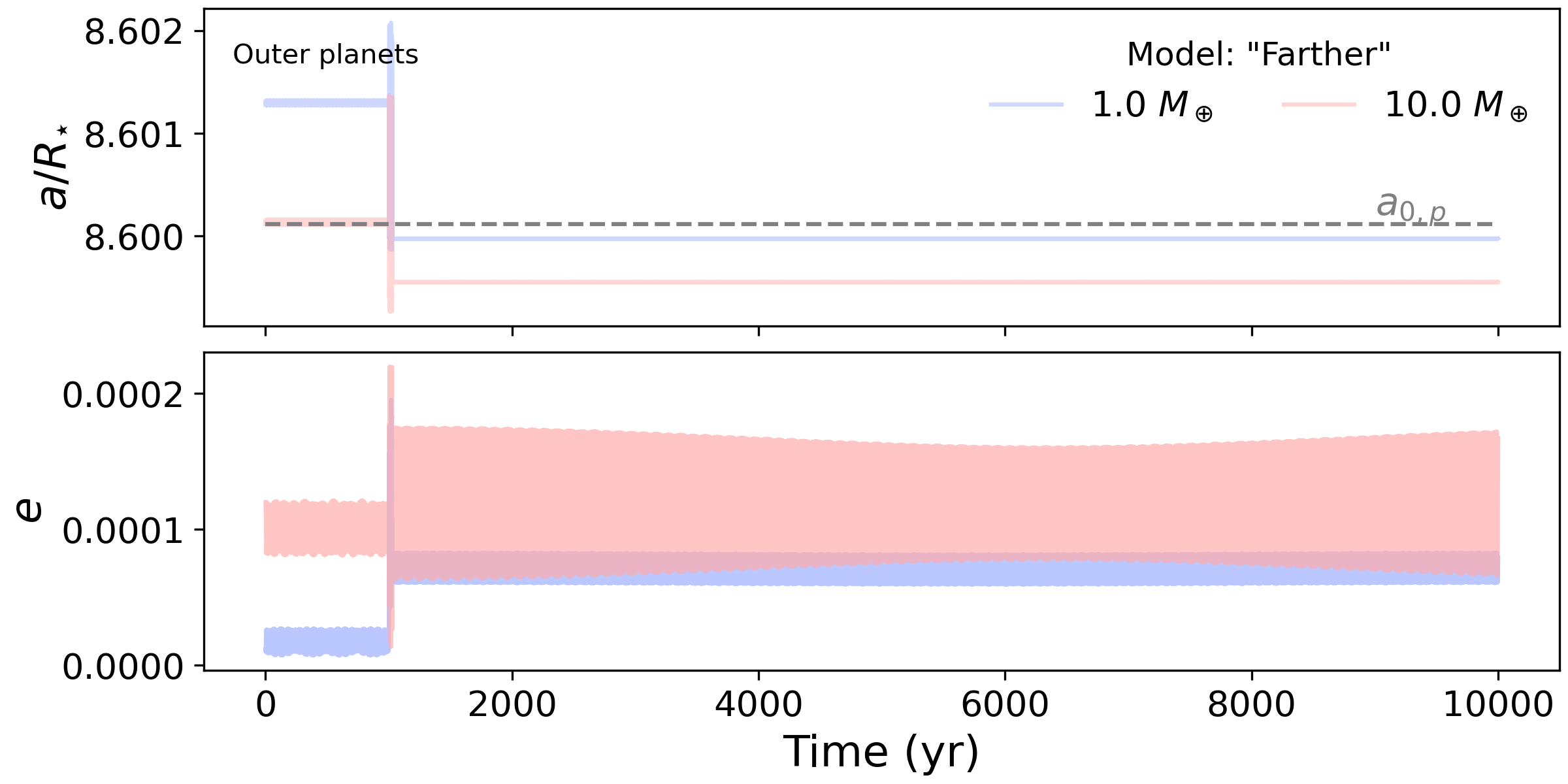}
  \includegraphics[width=0.99\columnwidth]{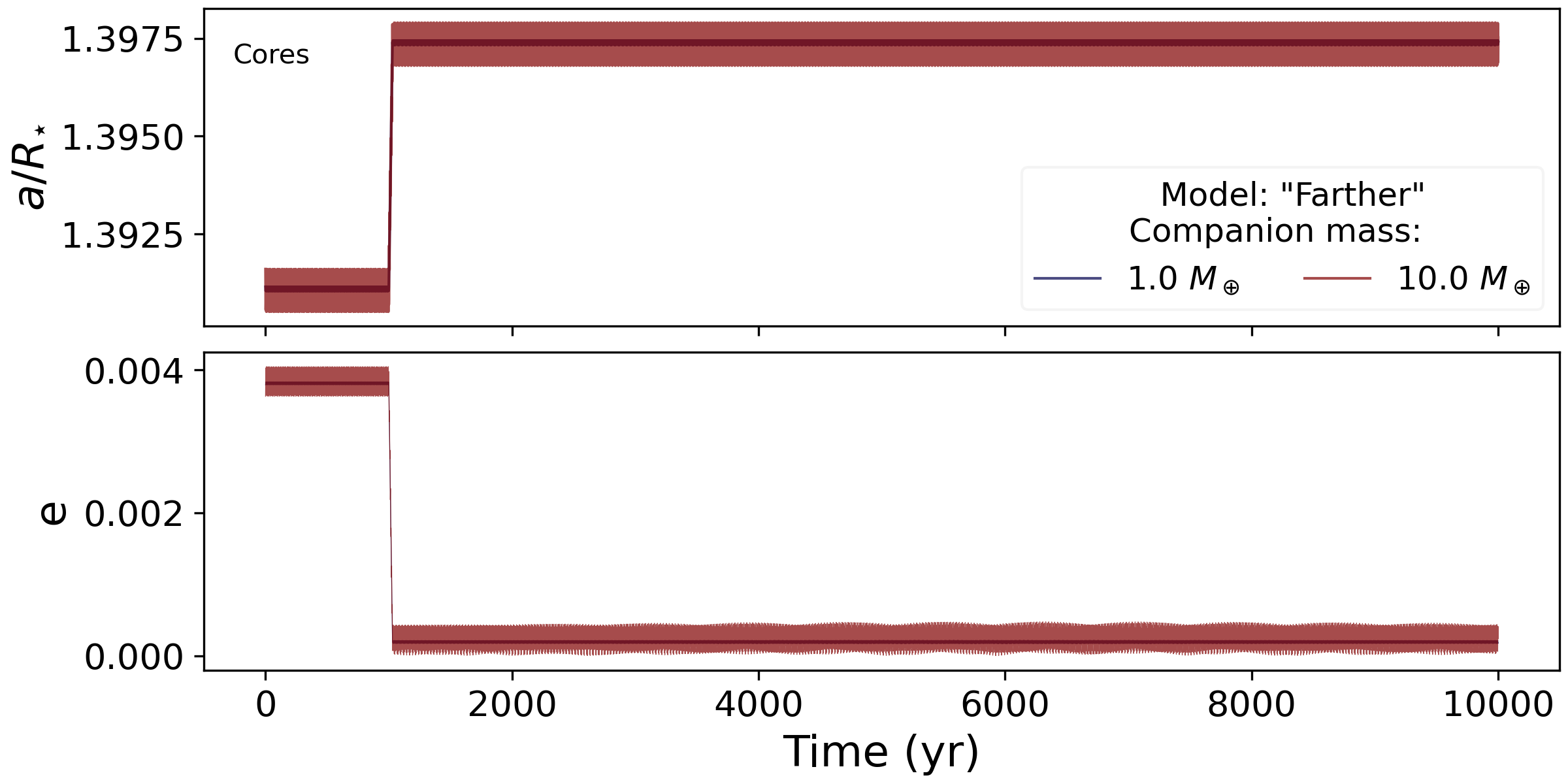}
  \caption{Orbital evolution during tidal disruption and mass transfer. Top two panels: Semimajor axis and eccentricity evolution of outer planets (1 and 10~$M_\oplus$) in the Closer regime. Bottom two panels: Evolution of the remnant core in the Farther regime. Dashed lines mark initial semimajor axes. Eccentricities rise from zero up to $\sim 0.07$ in chaotic configurations.}
\end{figure*}

We then modelled the dynamical consequences of tidal disruption by imposing a linear decrease in the giant planet’s mass—from its initial value to 5\% $M_\mathrm{Jup}$—over a 20-year interval starting at $t = 1000$\,yr, mimicking Roche-lobe overflow and mass transfer.

Figure~\ref{fig:plan2dinevol} shows the semimajor axis and eccentricity evolution during the disruption phase. Notably, eccentricities, initially zero, were excited up to $\sim 0.07$ in chaotic regions, particularly for the low-mass, close-in planets. In contrast, outer planets in the Farther regime remained dynamically stable, experiencing only modest eccentricity growth and slight inward migration. These results validate their role as plausible volatile-rich planets without requiring strong dynamical excitation.

While this work decouples hydrodynamics and gravitational dynamics, future fully self-consistent simulations—including gas flow, mass loss, and $N$-body interactions—could reveal subtler effects such as resonant migration, dynamical friction, and long-term eccentricity or inclination evolution.

\section{Detection threshold} \label{ap:detection}

The additional transit depth produced by a captured atmosphere is approximately given by
\begin{equation}
\Delta F \approx \dfrac{2 R_p A_H}{R_*^2},
\end{equation}
where $A_H$ is the effective height at which the atmosphere becomes optically thick in the tangential direction.

For an isothermal atmosphere with density profile $\rho(r) = \rho_0 \exp[-(r - R_p)/H]$, the height at which the optical depth reaches unity can be estimated as
\begin{equation}
A_H \approx H \log\left( \kappa_R \rho_0 \sqrt{2\pi R_p H} \right),
\end{equation}
where $\rho_0$ is the gas density at the base of the atmosphere, typically evaluated at the planetary radius $R_p$.

Relating the base density to the total atmospheric mass via $M_{\rm atm} \approx 4\pi R_p^2 H \rho_0$, we obtain
\begin{equation}
A_H \approx H \log\left( \dfrac{ \kappa_R M_{\rm atm} \sqrt{2\pi R_p H} }{ 4\pi R_p^2 } \right),
\end{equation}
which expresses $A_H$ as a function of atmospheric mass and opacity.

Substituting into the expression for $\Delta F$ yields
\begin{equation}\label{eq:deltaF}
\Delta F \approx \dfrac{2 R_p H}{R_*^2} \log\left( \dfrac{ \kappa_R M_{\rm atm} \sqrt{2\pi R_p H} }{ 4\pi R_p^2 } \right),
\end{equation}
providing the dependence of the transit depth on the captured mass and planetary properties. This formulation is valid in the optically thin regime, where the atmosphere does not saturate the transit signal.

The scale height, $H$, is set by hydrostatic equilibrium:
\begin{equation}
H = \dfrac{k_B T}{\mu m_p g},
\end{equation}
where $T$ is the atmospheric temperature, $\mu$ the mean molecular weight, $m_p$ the proton mass, and $g = G M_p / R_p^2$ the surface gravity.

\begin{figure}
    \centering
    \includegraphics[width=\columnwidth]{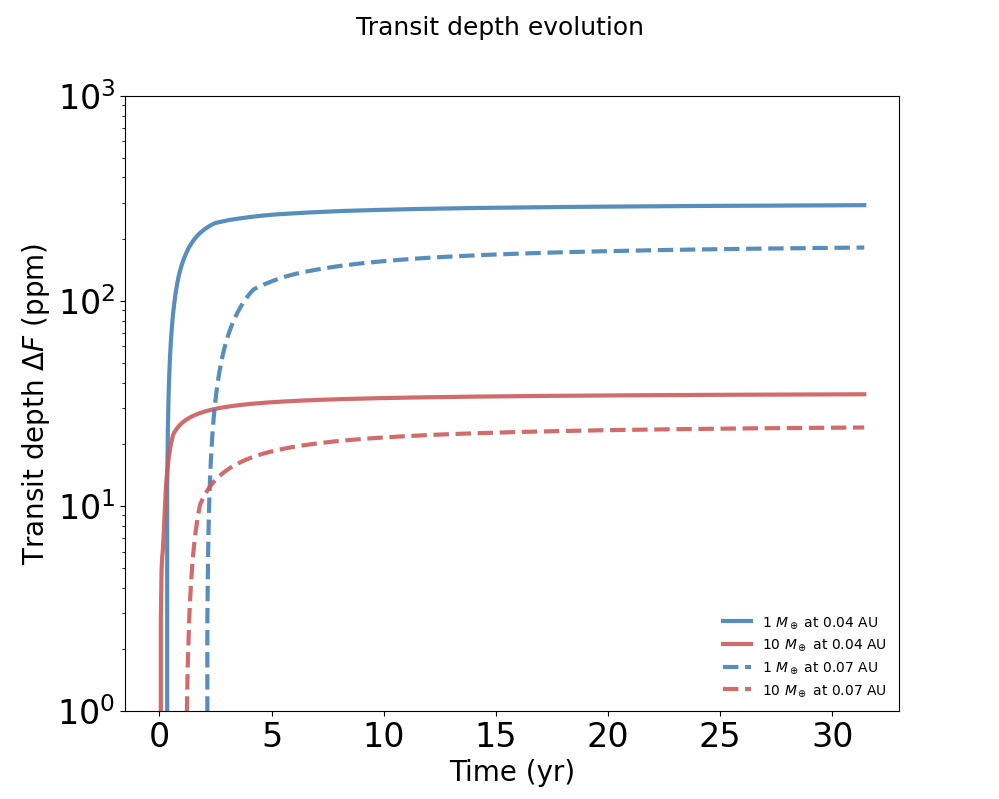}
    \caption{Transit depth evolution ($\Delta F$) as a function of time for two models with planetary cores of $1\,M_\oplus$ (dashed blue) and $10\,M_\oplus$ (solid red), both located at 0.07 au from a Solar-type star. The curves show the expected signal in ppm due to Kelvin–Helmholtz contraction of a captured H/He envelope under radiative cooling. After $\sim$30 years, the low-mass planet reaches $\Delta F \approx 272$ ppm, while the higher-mass planet yields $\Delta F \approx 28$ ppm, owing to its stronger surface gravity compressing the envelope.}
    \label{fig:transit}
\end{figure}

Using the equilibrium temperature and envelope masses derived from our simulations, we estimate the resulting transit signals (Eq.~\ref{eq:deltaF}). Figure~\ref{fig:transit} shows the time evolution of $\Delta F$ across models. The transit depth rises as the envelope grows via accretion, reaching $\sim$182–293 ppm for $1\,M_\oplus$ planets and $\sim$24–35 ppm for $10\,M_\oplus$ planets. This mass-dependent difference arises primarily from the stronger gravity of more massive cores, which compresses the envelope and reduces the atmospheric scale height.

Although the gas disc disperses rapidly on a viscous timescale of $\sim 10^2$ yr, the captured atmospheres persist much longer. The escape timescales are $\tau_{\rm esc} \sim 10^6$–$10^{8}$ yr depending on planet mass (see Sects.~\ref{subsec:Persistence} and \ref{subsec:DiscDispersal}).

\end{document}